\documentclass[11pt]{article}

\usepackage{epsfig}
\usepackage{subfig}
\usepackage{graphicx}
\usepackage{epstopdf}
\usepackage{amsfonts}
\usepackage{amssymb}
\usepackage{amsthm}
\usepackage{amsmath}
\usepackage{multirow}
\usepackage{color}
\usepackage{cite}
\usepackage{physics}
 \usepackage{makecell}
\def\beq{\begin{equation}}
\def\eeq{\end{equation}}
\def\bea{\begin{eqnarray}}
\def\eea{\end{eqnarray}}
\newcommand{\beqs}{\begin{subequations}}
\newcommand{\eeqs}{\end{subequations}}

\newcommand{\cref}[1]{Ref.~\cite{#1}}

\newcommand{\hh}{{\ensuremath{I{\kern-2.6pt h}}}}
\newcommand{\bhh}{{\ensuremath{\bar{I{\kern-2.6pt h}}}}}

\usepackage[colorlinks=true,allcolors=blue]{hyperref}

\begin{document}

\begin{titlepage}
	
\begin{center}
{\Large {\bf Kinetic Mixing, Proton Decay and Gravitational Waves in SO(10)}
}
\\[12mm]
Rinku Maji,$^{1}$
Qaisar Shafi$^{2}$
\end{center}
\vspace*{0.50cm}
	\centerline{$^{1}$ \it
		Cosmology, Gravity and Astroparticle Physics Group, Center for Theoretical Physics of the Universe,}
		\centerline{\it  Institute for Basic Science, Daejeon 34126, Republic of Korea}
	\vspace*{0.2cm}
	\centerline{$^{2}$ \it
		Bartol Research Institute, Department of Physics and 
		Astronomy,}
	\centerline{\it
		 University of Delaware, Newark, DE 19716, USA}
	\vspace*{1.20cm}

\begin{abstract}
\noindent
We present an $SO(10)$ model in which a dimension five operator induces kinetic mixing at the GUT scale between the abelian subgroups $U(1)_{B-L}$ and $U(1)_R$. We discuss in this framework gauge coupling unification and proton decay, as well as the appearance of superheavy quasistable strings with $G\mu \sim 10^{-8} - 10^{-5}$, where $\mu$ denotes the dimensionless string tension parameter. We use Bayesian analysis to show that for $G\mu$ values $\sim 4 \times 10^{-7} - 10^{-5}$, the gravitational wave spectrum emitted from the quasistable strings is in good agreement with the recent pulsar timing array data. Corresponding to $G \mu$ values $\sim 10^{-8} - 2 \times 10^{-7}$, proton decay is expected to occur at a rate accessible in the Hyper-Kamiokande experiment. Finally, we present the gravitational wave spectrum emitted by effectively stable strings with $G\mu\approx 10^{-8}$ that have experienced a certain amount of inflation. This can be tested with future detectors in the $\mu$Hz frequency range.
\end{abstract}

\end{titlepage}
\section{Introduction}
The quest for realistic grand unification based on $SO(10)$ \cite{Georgi:1974my, Fritzsch:1974nn} predicts topologically stable magnetic monopoles \cite{Lazarides:1980cc} and proton decay \cite{Buras:1977yy, Langacker:1980js}, and they are yet to be observed \cite{Ambrosio:2002qq, super-kamiokande:2020wjk}. In addition, topologically stable strings \cite{Kibble:1982ae}, and composite structures such as walls bounded by strings (WBS) \cite{Kibble:1982dd,Lazarides:2023iim, Maji:2023fba} and monopoles connected by strings \cite{Lazarides:2019xai} appear in $SO(10)$ depending on the symmetry breaking patterns.  The latter composite structures can give rise to the formation of the quasistable string (QSS) network \cite{martin:1996ea, Lazarides:2022jgr}. The recent NANOGrav 15 year data \cite{NANOGrav:2023gor, NANOGrav:2023hfp, NANOGrav:2023hvm} and other pulsar timing array (PTA) data \cite{EPTA:2023fyk,EPTA:2023xxk, Reardon:2023gzh, Xu_2023} impose a lower bound of $G \mu \lesssim 10^{-10}$ on topologically stable cosmic strings, where $\mu$ denotes the mass per unit length of the string. However, the PTA data is compatible with the gravitational wave background emitted by QSS and WBS \cite{Maji:2023fba, Lazarides:2023ksx} networks with superheavy strings having $G \mu \sim 10^{-6}$. Another example is based on the so-called metastable strings (MSS) \cite{Buchmuller:2019gfy,Buchmuller:2021mbb}, where the topologically unstable strings decay via quantum mechanical tunneling. For some recent studies of MSS based on grand unification see Refs.~\cite{Antusch:2023zjk,Lazarides:2023rqf,Maji:2023fhv, Ahmed:2023rky,afzal:2023kqs,Antusch:2024nqg,Maji:2024pll}. Stable string networks with a reduced intercommutation probability ($\mathcal{P}$) such as cosmic superstrings are expected to generate stochastic gravitational waves enhanced by a factor $\sim 1/\mathcal{P}$ \cite{Dvali:2003zj, Jackson:2004zg, Sakellariadou:2004wq, Avgoustidis:2005nv, Blanco-Pillado:2017rnf, Ellis:2020ena, Yamada:2022aax,Yamada:2022imq}. In turn, they can explain the PTA data with $G\mu\sim 10^{-12}-10^{-11}$ and $\mathcal{P}\sim 10^{-3}-10^{-2}$ \cite{NANOGrav:2023hvm, Ellis:2023tsl, Yamada:2023thl}.

 Another motivation for cosmic strings arises from an early analysis of data from the Jame Webb Space Telescope (JWST) which indicates that the standard $\Lambda$CDM model may not adequately explain the mass fraction in early (high redshift) halos \cite{Labb__2023}. It has recently been proposed that cosmic strings with a string tension $G \mu \sim 10^{-8}$ could provide the perturbations necessary for generating the halos at high redshifts \cite{jiao:2023wcn,Wang:2023len}.

 In this paper, based on $SO(10)$, we include a kinetic mixing of the abelian gauge groups \cite{holdom:1985ag,delaguila:1988jz,delaguila:1995rb,bertolini:2009qj,deromeri:2011ie,fonseca:2013bua,chakrabortty:2020otp}, namely $U(1)_R\times U(1)_{B-L}$, which is induced by a dimension five operator suppressed by the cut-off scale. We explore the formation of superheavy QSS from the breaking of $SU(4)_c\times SU(2)_L\times SU(2)_R$ to $SU(3)_c\times U(1)_{B-L}\times SU(2)_L\times U(1)_R$. The $SU(4)_c\times SU(2)_L\times SU(2)_R$ symmetry \cite{Pati:1974yy} is a maximal subgroup of $SO(10)$ and its breaking produces topologically stable monopoles which carry two units of Dirac magnetic charge after the electroweak symmetry breaking \cite{Lazarides:2019xai}. The mixing introduced by the higher dimensional operator is crucial to obtain the near-GUT scale strings. We perform a Bayesian analysis to show that the stochastic gravitational wave background emitted from the QSS network is compatible with the PTA data. We also show how effectively stable cosmic strings with $G\mu\sim 10^{-8}$ that experience an appropriate number of $e$-foldings during inflation generate a gravitational wave spectrum that can be tested in the $\mu$Hz frequency range. For a suitable range of $G\mu$ values we show that proton decay should be accessible in the Hyper-Kamiokande experiment \cite{dealtry:2019ldr}.
 
The paper is arranged as follows. In Section~\ref{sec:2} we describe the $SO(10)$ symmetry breaking pattern and identify the red and blue monopoles associated with the quasistable cosmic strings. In Section~\ref{sec:KM} we show how the kinetic mixing between the abelian groups $U(1)_{B-L}$ and $U(1)_R$ emerges in this model in the presence of a dimension five operator. Another dimension five operator responsible for generating the Majorana masses for the right handed neutrinos is briefly mentioned. In Section~\ref{sec:unific-sol} we discuss gauge coupling unification and proton decay, and show that for a range of $G\mu$ values, proton decay should be accessible at the Hyper-Kamiokande experiment. In Section \ref{sec:5-GWs-QSS} we compare the gravitational wave background emitted by quasistable strings with the PTA measurements. Section \ref{sec:6-GWs-JWST} shows the spectrum from essentially stable strings with $G \mu\approx 10^{-8}$ that have experienced a significant amount of inflation. This will be tested in future experiments operating in the $\mu$Hz frequency range. It has been suggested \cite{jiao:2023wcn} that cosmic strings with $G \mu$ values of this magnitude may provide seeds for high redshift galaxies observed by the James Webb Space Telescope. Our conclusions are summarized in Section \ref{sec:summary}.

\section{SO(10) symmetry breaking, monopoles and strings}
\label{sec:2}
We consider the following $SO(10)$ symmetry breaking to the SM:
\begin{align}\label{eq:breaking-chain}
SO(10) &\xrightarrow[M_U]{\left<210\right>} SU(4)_c \times SU(2)_L \times SU(2)_R \nonumber\\
& \xrightarrow[M_U]{\left<(15,1,3)\in 210\right>} SU(3)_c \times U(1)_{B-L} \times SU(2)_L \times U(1)_R \nonumber\\
& \xrightarrow[M_R]{\left<(1,1,1,-\frac{1}{2})\in 16\right>} SU(3)_c \times SU(2)_L \times U(1)_Y.
\end{align}    
The symmetry breakings via the VEVs of the scalar multiplet $210$ produce magnetic monopoles at the GUT scale $M_U$. The breaking of $SO(10)$ to $SU(4)_c\times SU(2)_L\times SU(2)_R$ (see Ref.~\cite{slansky:1981yr} for a review) via the VEV of $(1,1,1)\in 210$ produces topologically stable GUT monopoles which transform into Dirac monopoles carrying one unit of Dirac magnetic charge \cite{Dirac:1931kp} after the electroweak symmetry breaking \cite{Lazarides:1980cc,Lazarides:2019xai}. The symmetry breakings $SU(4)_c\to SU(3)_c\times U(1)_{B-L}$ and $SU(2)_R\to U(1)_R$ by the adjoint scalar $(15,1,3)\in 210$ produce monopoles which we call `red' and `blue' monopoles respectively, as in Ref.~\cite{Lazarides:2019xai}. The subsequent breaking of $U(1)_{B-L}\times U(1)_R$ to $U(1)_Y$ is achieved via the VEV of the SM singlet component in a 16-dimensional scalar field. This breaking produces cosmic strings which are not topologically stable. These strings connect \cite{Lazarides:2019xai,Lazarides:2023iim,Lazarides:2024niy}:
\begin{enumerate}
\item[1.] a blue monopole to a red monopole,
\item[2.] a monopole to its anti-monopole for both red and blue monopoles,
\item[3.] ends on itself forming a loop.
\end{enumerate}
These strings can decay via quantum tunneling of monopole-antimonopoles pairs with a decay width per unit length given by \cite{preskill:1992ck}
\begin{align}
\Gamma_d = \frac{\mu}{2\pi}\exp\left(-\pi m_M^2/\mu\right) ,
\end{align}
where $m_M$ is the mass of the monopole and $\mu$ is the string tension. The strings are effectively stable against the breaking via Schwinger pair-production unless the breaking scales $M_U$ and $M_R$, associated with the monopoles and strings respectively, are almost degenerate.

We expect that there will be one red and one blue monopole at production within a correlated region (causally connected) in space whose length is approximately given by $min[H^{-1},m_{\rm eff}^{-1}]$. Here, $H$ denotes the Hubble parameter and $m_{\rm eff}$ is the effective mass of the scalar during the phase transition at the scale $M_U$. These monopoles are stable until they get connected by cosmic strings.

The blue and red monopoles connected by strings combine to form stable magnetic monopoles with two unit of Dirac magnetic charge. Therefore, the monopoles must encounter sufficient number of $e$-foldings ($N_M$) to satisfy the current bound on the monopole flux from experiments such as MACRO \cite{Ambrosio:2002qq}, IceCube \cite{IceCube:2021eye} and ANTARES \cite{ANTARES:2022zbr}. On the other hand, the monopole-antimonopole pairs connected by strings annihilate. Since the monopoles undergo some $e$-foldings their separation becomes larger than the particle horizon. In this situation the strings connecting monopoles perform a random walk with a step comparable to the size of the particle horizon, behave as a network of stable strings, inter-commute, form loops and radiate gravitational waves. We assume that the monopoles are partially inflated ($N_M\lesssim 50$). This yields the scenario of `quasistable' cosmic strings, as discussed in Ref.~\cite{Lazarides:2022jgr}, when the $\nu^c$-type component in $16$ dimensional scalar acquires the VEV at scale $M_R$. 
\section{Dimension five operators and kinetic mixing of $U(1)_{B-L}\times U(1)_R$}
\label{sec:KM}
The intermediate symmetry $SU(3)_c \times U(1)_{B-L} \times SU(2)_L \times U(1)_R$ allows a mixing term in the gauge kinetic part of the Lagrangian given by
\begin{align}
\label{eq:km}
-\frac{1}{4}F^{X}_{\mu\nu}F^{X\mu\nu}-\frac{1}{4}F^{R}_{\mu\nu}F^{R\mu\nu}-\frac{\epsilon}{2}F^{X}_{\mu\nu}F^{R\mu\nu},
\end{align}
where $X\equiv B-L$, $F^{p}_{\mu\nu}=\partial_\mu A^p_\nu-\partial_\nu A^p_\mu$  ($p=X,R$) are the field strength tensors, and $\epsilon$ is the mixing parameter. The parent symmetries ($SO(10)$ and $SU(4)_c\times SU(2)_L\times SU(2)_R$) do not allow this mixing term at the renormalizable level. We consider the following dimension five operators in $SO(10)$ :
\begin{align}\label{eq:dim-5-gut}
\frac{\mathcal{C}_1}{\Lambda}G_{\mu\nu}{210}_{H}G^{\mu\nu},\quad \frac{\mathcal{C}_2}{\Lambda}16_F{16}_H{16}_H 16_F, \quad \frac{\mathcal{C}_3}{\Lambda}16_F \overline{16}_H\overline{16}_H 16_F,\quad \frac{\mathcal{C}_4}{\Lambda}16_F{10}_H{210}_H 16_F
\end{align}
where $\mathcal{C}_n$ ($n=1,2,3,4$) are dimensionless couplings and $\Lambda$ is the UV cut-off scale. Following the weak gravity conjecture \cite{Arkani-Hamed:2006emk}, we expect $\Lambda \sim \sqrt{\alpha_U} m_{\rm Pl}$, where $\alpha_U$ denotes the unified gauge coupling parameter and $m_{\rm Pl} = 2.4 \times 10^{18}$ GeV is the reduced Planck scale.
 The first operator contains a mixing term given by
\begin{align}
G^{a\mu\nu}\Phi_{ai}G^i_{\mu\nu}
\end{align} 
where $\Phi_{ai}\equiv (15,1,3)\in 210_H$. The SM singlet in $\Phi_{ai}$ achieves a VEV at the GUT scale and generates the kinetic mixing term in Eq.~\eqref{eq:km}. Therefore, $\epsilon\sim \mathcal{O}(M_U/\Lambda)$ at the GUT scale.

We transform the kinetic terms given in Eq.~\eqref{eq:km} into the canonical form through a $GL(2,\mathbb{R})$ transformation of the gauge fields ($A_\mu^X,A_\mu^R$) given by \cite{Babu:1997st,Brahmachari:2014aya}
\begin{align}
\label{eq:gl2R}
\begin{pmatrix}
A_\mu^X \\
A_\mu^R
\end{pmatrix}
=
\begin{pmatrix}
1 & \frac{-\epsilon}{\sqrt{1-\epsilon^2}} \\
0 & \frac{1}{\sqrt{1-\epsilon^2}}
\end{pmatrix}
\begin{pmatrix}
B_\mu^X \\
B_\mu^R
\end{pmatrix}.
\end{align}
However, this transforms the diagonal gauge coupling $diag[g_X,g_R]$ interactions in the covariant derivative part to an upper triangular matrix form given by
\begin{align}
\label{eq:G}
G = \begin{pmatrix}
g_{XX} & g_{XR} \\
0 & g_{RR}
\end{pmatrix},
\end{align}
with 
\begin{align}
g_{XX} = g_X, \quad
g_{XR} = -\frac{\epsilon g_X}{\sqrt{1-\epsilon^2}}, \quad
g_{RR}=\frac{g_R}{\sqrt{1-\epsilon^2}}.
\end{align}
At the GUT scale $M_U$ we set $g_X=g_R\equiv g_U$. We have the matching condition at the breaking scale $M_R$ given by
\begin{align}
\frac{1}{\alpha_Y(M_R)}=4\pi P(GG^T)^{-1}P^T,
\end{align}
with $P=(\sqrt{\frac{2}{5}},\sqrt{\frac{3}{5}})$, which yields 
\begin{align}
{g_Y^2}=\frac{5 g_{XX}^2 g_{RR}^2}{3 g_{XX}^2-2 \sqrt{6} g_{XX} g_{XR}+2 \left(g_{XR}^2+g_{RR}^2\right)}  \; .
\end{align} 

We use the package PyR@TE \cite{Sartore:2020gou} to compute the renormalization group evolution of the gauge couplings with the non-diagonal $G$ in Eq.~\eqref{eq:G}. The RGE equation for a gauge coupling $g_i$ is  given by \cite{Georgi:1974yf,jones:1981we}
\begin{align}
\mu \frac{dg_i}{d\mu}=\beta_i,
\end{align}
  where $\mu$ is the renormalization scale and $\beta_i$ is the $\beta$-function. The $\beta$-functions for the gauge couplings at the intermediate symmetry $SU(3)_c\times U(1)_{B-L}\times SU(2)_L\times U(1)_R$ are given in Appendix~\ref{appen:A}. 
  
 Before closing this section a brief description of the Yukawa couplings are in order. The third dimension five operator in Eq.~\ref{eq:dim-5-gut} provides a Majorana mass to the right-handed neutrino \cite{lazarides:1980nt, Babu:1992ia,Babu:1998wi}, and the second and third terms can provide Dirac mass terms. The standard model Higgs doublet will arise from the linear combination of the $SU(2)_L$ doublets in a 10-dimensional complex scalar with the Yukawa couplings 
\begin{align}
y_116_F16_F10_H + y_216_F16_F10_H^* + \mathrm{H.c.} ,
\end{align}
and the lepton-like doublet in the $16_H$. For a recent study see Ref.~\cite{Shukla:2024owy}.

\section{Unification and proton decay}
\label{sec:unific-sol}
We utilize the $\chi^2$ statistics to find the solution compatible with the unification given as \cite{chakrabortty:2017mgi,chakrabortty:2019fov}
\begin{align}
\chi^2=\sum_{i=1}^3\frac{(g_{i,\rm th}^2-g_{i,\rm exp}^2)^2}{\sigma_{i,\rm exp}^2},
\end{align} 
where $g_i$ denotes the SM gauge couplings whose experimental values ($g_{i,\rm exp}^2\pm \sigma_{i,\rm exp}$) can be obtained from the electroweak observables \cite{ParticleDataGroup:2022pth}. For a given intermediate scale $M_R$ we use the RGE to compute the theoretical predictions of the gauge couplings ($g_{i,\rm th}^2$) as a function of the unification scale $M_U$, the unified gauge coupling $g_U$, and the off-diagonal coupling $g_{XR}$ at $M_U$. The unification solutions are obtained after the $\chi^2$ minimization.

\begin{figure}[h!]
\begin{center}
\includegraphics[width=0.7\textwidth]{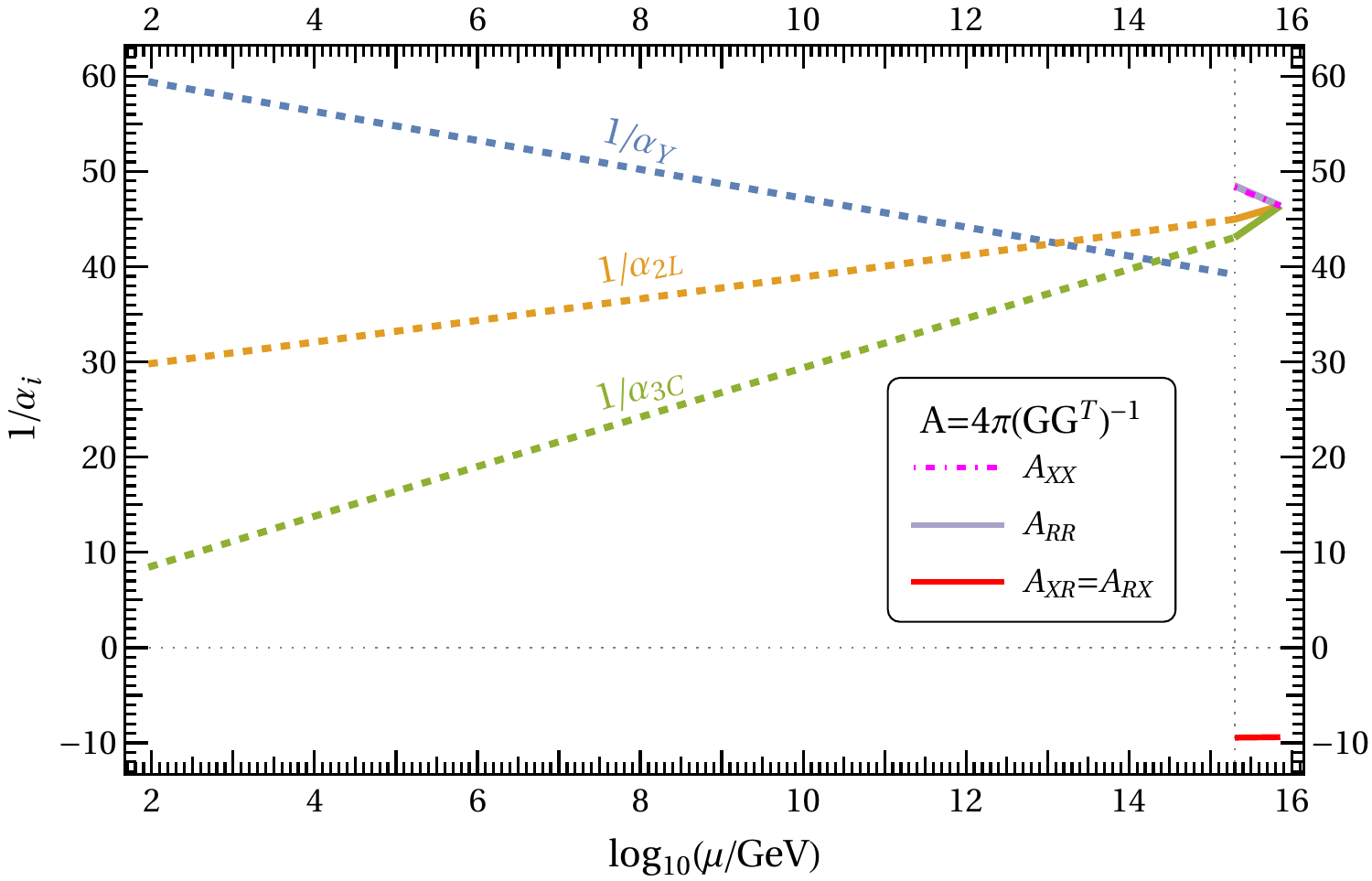}
\caption{RGE running of the gauge couplings $1/\alpha_i = 4\pi/g_i^2$ (where, $i=Y, 2L, 3C$ for the SM) for successful unification with $M_R=2\times 10^{15}$ GeV. The elements of gauge coupling matrix $A\equiv 4\pi (GG^T)^{-1}$ for abelian mixing is obtained from Eq.~\eqref{eq:G} and $1/\alpha_i = A_i$ with $i=XX, XR, RX, RR$ in this case.}
\label{fig:runplt}
\end{center}
\end{figure}
\begin{figure}[h!]
\begin{center}
\includegraphics[width=0.8\textwidth]{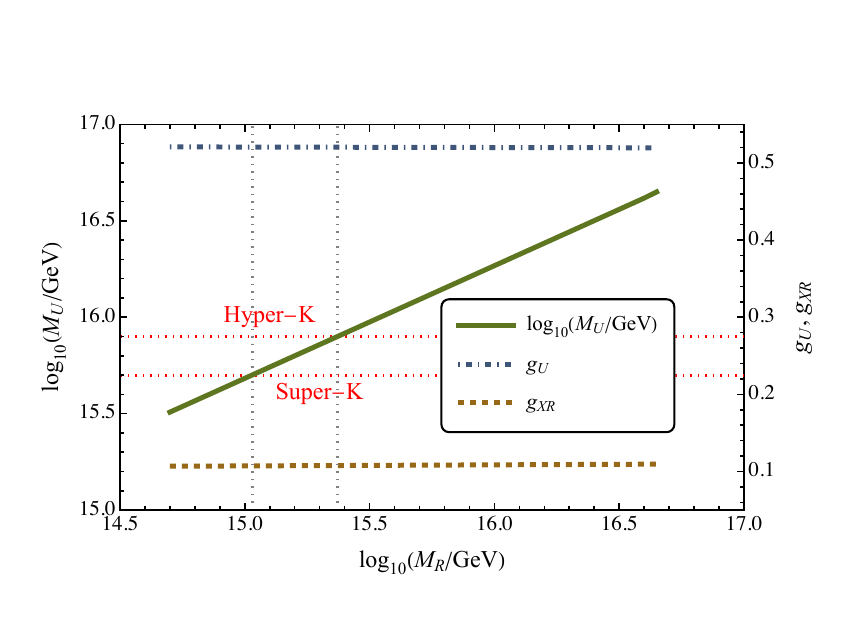}
\caption{Unification scale $M_U$ on left $y$ axis, and unified gauge couplings $g_U$ and off-diagonal coupling $g_{XR}$ at $M_U$ on right $y$ axis as a function of the intermediate scale $M_R$ for the symmetry breaking chain in Eq.~\eqref{eq:breaking-chain}.}
\label{fig:g-sol}
\end{center}
\end{figure}
 Fig.~\ref{fig:runplt} depicts the running of the gauge couplings ($1/\alpha_i = 4\pi/g_i^2$ or the components in $4\pi (GG^T)^{-1}$ for the Abelian mixing) for a successful unification with an intermediate scale $M_R=2\times 10^{15}$ GeV. Fig.~\ref{fig:g-sol} shows the unification scale $M_U$, the unified gauge coupling $g_U$, and the off-diagonal coupling $g_{XR}$ at $M_U$ as a function of the intermediate scale $M_R$ for successful unification. The solution region with successful unification are $\log_{10}(M_U/\mathrm{GeV})=[15.5,16.6]$ and $\log_{10}(M_R/\mathrm{GeV})=[14.7,16.6]$ with slowly increasing values around $g_U\simeq 0.52$ and $g_{XR}\simeq 0.11$. The solutions for $\log_{10}(M_U/\mathrm{GeV})< 15.7$ associated with $\log_{10}(M_R/\mathrm{GeV})<15.0$ are excluded by the Super-Kamiokande experiment \cite{super-kamiokande:2020wjk}. The solution region $\log_{10}(M_U/\mathrm{GeV})=[15.7,15.9]$ for $\log_{10}(M_R/\mathrm{GeV}) = [15.0,15.4]$ will be probed in the future data from the Hyper-Kamiokande experiment \cite{dealtry:2019ldr}. Recall that there is one intermediate scale between the electroweak and GUT scale in our model. It is worth mentioning that the solution region will expand if we include an additional intermediate scale. There will be modifications in the gauge coupling matching condition from integrating out the heavy degrees of freedom \cite{weinberg198051,hall:1980kf} and from the dimension five operator \cite{Hill:1983xh, Shafi:1983gz,Hall:1992kq, Calmet:2008df, chakrabortty:2008zk,chakrabortty:2009xm,calmet:2009hp,chakrabortty:2010xq,calmet:2011ic,chakrabortty:2017mgi,Lazarides:2022ezc}.
 
\section{Gravitational waves from quasistable strings}  
\label{sec:5-GWs-QSS}
\begin{figure}[h!]
\begin{center}
\includegraphics[width=0.7\textwidth]{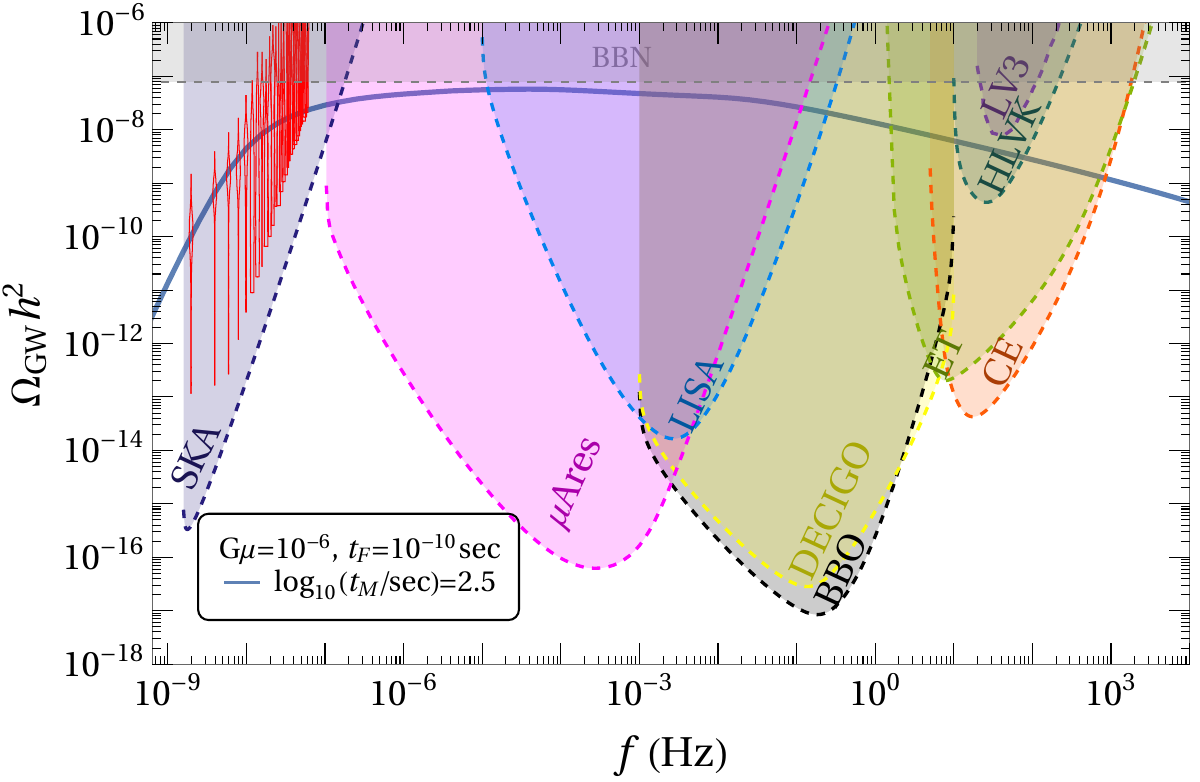}
\end{center}
\caption{Gravitational wave background from quasistable cosmic strings with $G\mu = 10^{-6}$ and $t_M = 10^{2.5}$ sec.  The red violins depict the posteriors of an HD correlated free spectral reconstruction of the NANOGrav 15 year data. We assume the string network enters a scaling regime at the time $t_F=10^{-10}$ sec after their horizon re-entry in order to satisfy the LIGO-VIRGO third run data \cite{ligoscientific:2021nrg}. The gray region depicts the bound from Big Bang Nucleosynthesis (BBN) \cite{Mangano:2011ar}. We have shown the power-law integrated sensitivity curves \cite{Thrane:2013oya, Schmitz:2020syl} for planned experiments, namely, HLVK \cite{KAGRA:2013rdx}, CE \cite{Regimbau:2016ike}, ET \cite{Mentasti:2020yyd}, DECIGO \cite{Sato_2017}, BBO \cite{Crowder:2005nr, Corbin:2005ny}, LISA \cite{Bartolo:2016ami, amaroseoane2017laser}, $\mu$Ares \cite{Sesana:2019vho} and SKA \cite{5136190, Janssen:2014dka}.}\label{fig:GWs-QSS}
\end{figure}
The symmetry breaking at $M_R$ produces cosmic strings, which are effectively stable against the monopole-antimonopole pair production. The dimensionless string tension parameter falls within the range $G\mu \sim [10^{-8},10^{-5}]$ for the unification solution. A network of quasistable strings (QSS) is formed if the monopoles are partially inflated. Let us recall that cosmic strings are metastable if they break up via the quantum mechanical tunneling of monopole-antimonopole pairs. In contrast, quasistable strings (QSS) break up through the attachment of monopole-antimonopole pairs that reenter the horizon after experiencing a certain amount of inflation. Quantum tunneling in the QSS scenario is exponentially suppressed.
\begin{figure}[h!]
\begin{center}
\includegraphics[width=0.7\linewidth]{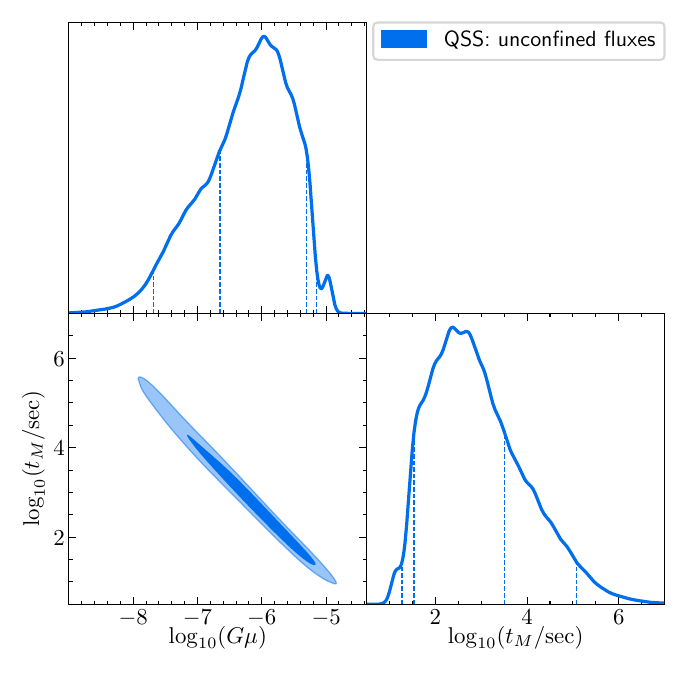}
\caption{Triangular plot of the posterior distribution of $G\mu$ and $t_M$ for QSS with unconfined fluxes. The off-diagonal plot shows $1\sigma$ (dark) and $2\sigma$ (light) Bayesian credible regions. The diagonal plots dictate the marginalized 1D distributions with the dashed vertical lines indicating the $68\%$ and $95\%$ credible intervals.}\label{fig:triang-QSS}
\end{center}
\end{figure}

   The monopoles connected by strings carry unconfined flux and decay by radiating massless gauge bosons \cite{Berezinsky:1997kd, Leblond:2009fq, Kibble:2015twa,Lazarides:2024niy} as they oscillate after their horizon re-entry. Therefore, the contributions from string loops dominate the stochastic gravitational waves background. We integrate the unresolvable bursts with a waveform $h(f,l,z)$ from the redshift $z_*$ to $z_F$ to compute the gravitational wave background given by
  \begin{align}\label{eq:GWs}
\Omega_{\rm GW} (f)= \frac{4\pi^2}{3H_0^2}f^3\int_{z_{*}}^{z_F} dz \int_0^{d_H} dl \, h^2(f,l,z)\frac{d^2R}{dz \, dl} \ ,
\end{align}
where $\frac{d^2R}{dz \, dl}$ is the burst rate coming from a loop distribution $n(l,t)$, $d_H$ denotes the particle horizon at time $t(z)$, and $H_0$ is the Hubble parameter at the present era. Extensive studies on the gravitational waves from strings can be found in the literature \cite{Vilenkin:1981kz, Vachaspati:1984gt,Kibble:1984hp,Vilenkin:2000jqa,Damour:2001bk,Vanchurin:2005pa,Ringeval:2005kr,Olum:2006ix,Leblond:2009fq, Olmez:2010bi,Blanco-Pillado:2013qja,Blanco-Pillado:2017oxo,Cui:2018rwi,Buchmuller:2021mbb,Roshan:2024qnv}.

\begin{table}[h!]
\centering
\begin{tabular}{lcccc}
\hline
\multirow{2}{*}{Model} & \multirow{2}{*}{Parameters} & \multicolumn{2}{c}{Credible Intervals} \\
\cline{3-4}
 & & 68\% & 95\% \\
\hline
\multirow{2}{*}{QSS} & $\log_{10}(G\mu)$ & $[-6.65, -5.31]$ & $[-7.68, -5.15]$ \\
 & $\log_{10}(t_M/\mathrm{sec})$ & $[1.53, 3.51]$ & $[1.27, 5.08]$ \\
\hline
\end{tabular}
\caption{Bayesian credible intervals for the model parameters.}
\label{tab:params}
\end{table}
  Figure \ref{fig:GWs-QSS} depicts the gravitational wave background emitted by the quasistable cosmic strings with $G\mu = 10^{-6}$ and a monopole re-entry time $t_M = 10^{2.5}$ sec. In order to satisfy the advanced LIGO-VIRGO bound \cite{ligoscientific:2021nrg}, we assume that the string network enters a scaling regime at a late time $t_F=10^{-10}$ sec after their horizon re-entry. We utilize the wrapper PTArcade \cite{Mitridate:2023oar} to reconstruct the Bayesian credible region of the model parameters $G\mu$ and $t_M$ in QSS with the NANOGrav 15 year data. We employ the \texttt{Ceffyl} package \cite{Lamb:2023jls} with the Hellings-Downs (HD) correlation \cite{Hellings:1983fr} between the pulsars to obtain the posterior distribution. The triangular plot in Figure \ref{fig:triang-QSS} shows the posterior distribution of $G\mu$ and $t_M$. Table~\ref{tab:params} shows the 68\% and 95\% Bayesian credible intervals of the model parameters.

\section{Gravitational waves from effectively stable strings}
\label{sec:6-GWs-JWST}
In this section we discuss a scenario in which the cosmic strings are effectively stable. This is realized by inflating away the monopoles that otherwise would lead to QSS strings. We are interested in this case because effectively stable strings with $G \mu \sim 10^{-8}$ may have the potential to provide seeds for the high redshift galaxies \cite{jiao:2023wcn} suggested in the early JWST observations \cite{Labb__2023}. We recall that the recent PTA observations impose a significantly stringent constraint, $G \mu \lesssim 10^{-10}$, on stable cosmic strings. We aim to circumvent this constraint by requiring that the string network experiences a significant amount of inflation.
  
\begin{figure}[h!]
\begin{center}
\includegraphics[width=0.7\textwidth]{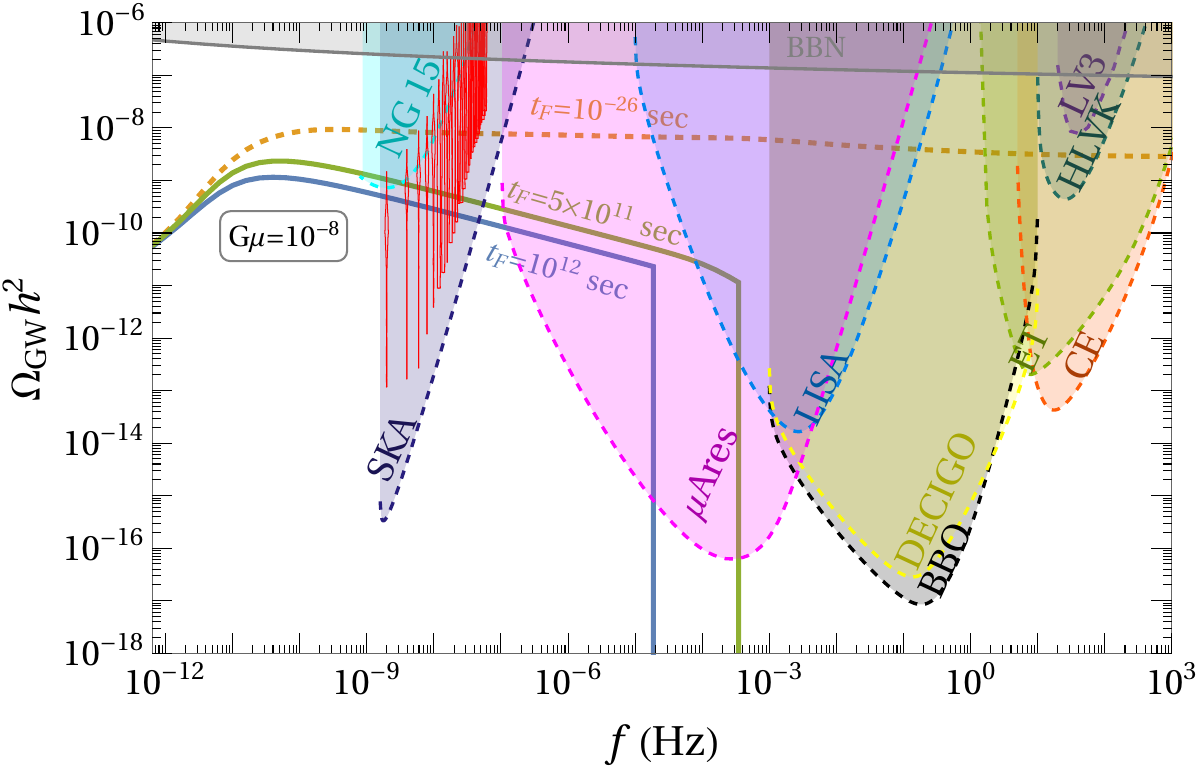}
\end{center}
\caption{Gravitational wave background from effectively stable cosmic strings with $G\mu = 10^{-8}$.  The NANOGrav 15 year data disfavors the stochastic background from stable strings with such high $G\mu$ values. Here, the strings experience a certain number of $e$-foldings and the string network enters a scaling regime at the time $t_F$ after their horizon re-entry. The gravitational waves become a subdominant component to the stochastic background to explain the strong evidence in NANOGrav 15 year data (red violins). The cyan region depicts the bound from NANOGrav 15 year data.}\label{fig:GWs-JWST}
\end{figure}
  The strings become effectively stable if the $SU(4)_c\times SU(2)_L\times SU(2)_R$ symmetry is never restored or the monopoles are completely inflated away. In this case, the data from CMB \cite{charnock:2016nzm,lizarraga:2016onn} and LIGO-VIRGO third run put a constraint $G\mu\lesssim 10^{-7}$. The PTA data provides a more stringent bound $G\mu\lesssim 10^{-10}$. Partial inflation of the strings helps to circumvent this bound or at least ameliorate the tension.
  
  Figure \ref{fig:GWs-JWST} shows the impact of inflation on the gravitational wave background for $G\mu\sim 10^{-8}$. The strings experience a certain amount of $e$-foldings, re-enter the horizon and the string network enters a scaling regime at a subsequent time $t_F$. For $t_F > 10^{11}$ sec, the gravitational waves from strings are suppressed and contribute a subdominant component to the gravitational wave background observed in the NANOGrav 15 year data. There will be a $f^{-1/3}$ dependence in the gravitational wave spectra which can be observed with a detector in the $\mu$Hz region \cite{Sesana:2019vho}.
 Finally, cosmic strings with $G\mu>10^{-7}$ should reenter the horizon after the CMB epoch. The stochastic gravitational wave background in this case is generated from the decay of loops that form and decay during the matter-dominated universe and is highly suppressed.
 
   Before concluding, it is worth noting that the minimum stellar mass $M_F$ seeded by the string loops at a redshift $z$ will depend on the time $t_F$. A comprehensive analysis of the interplay between inflation, gravitational waves and seeds for early galaxy formation is necessary which we leave for a future study.
\section{Summary}
\label{sec:summary}
We have presented a novel $SO(10)$ model in which a dimension five operator suppressed by a cut-off scale generates a kinetic mixing among the abelian symmetries $U(1)_{B-L}$ and $U(1)_R$, which plays an essential role in realizing gauge coupling unification. Proton decay events should be observed in the Hyper-Kamiokande experiment for string tension $G\mu \sim 10^{-8}-2\times 10^{-7}$. The superheavy cosmic strings are not topologically stable and form a quasistable string network with $G\mu\sim 10^{-8}-10^{-5}$. These quasistable strings can explain the pulsar timing array data for $G\mu\simeq 4\times 10^{-7}-10^{-5}$, that we verify using Bayesian analysis. Effectively stable string networks experience a sufficient number of $e$-foldings to satisfy the bounds on the gravitational waves from the PTA and advanced LIGO-VIRGO, as well as the CMB constraint. The gravitational waves from such an inflated string network with $G\mu\sim 10^{-8}$ can be accessible with a detector sensitive in the $\mu$Hz frequency range. We note that effectively stable strings with $G\mu \simeq 10^{-8}$ can generate the seeds for high redshift galaxies observed by the James Webb Space Telescope.
\section{Acknowledgment}
R.M. is supported by the Institute for Basic Science under the project code: IBS-R018-D3. We thank Joydeep Chakrabortty, Tripurari Srivastava and Amit Tiwari for helpful discussions.
\appendix
\section{Beta functions for $SU(3)_c \times U(1)_{B-L} \times SU(2)_L \times U(1)_R$}
\label{appen:A}

\begin{align}
 \beta_{3C} =& -\frac{7 g_{3C}^3}{16 \pi^2}+\frac{1}{256 \pi ^4}\left(\frac{3 g_{XR}^2 g_{3C}^3}{2}+\frac{9 g_{2L}^2 g_{3C}^3}{2}-26 g_{3C}^5+\frac{g_{3C}^3 g_{XX}^2}{2}+\frac{3 g_{3C}^3 g_{RR}^2}{2}\right), \\
 \beta_{2L} =& -\frac{3 g_{2L}^3}{16 \pi ^2}+\frac{1}{256 \pi ^4}\left(g_{XR}^2 g_{2L}^3+8 g_{2L}^5+12 g_{2L}^3 g_{3C}^2+\frac{3 g_{2L}^3 g_{XX}^2}{2}+g_{2L}^3 g_{RR}^2\right), \\
 \beta_{RR} =& \frac{53 g_{RR}^3}{192 \pi ^2}+\frac{1}{256 \pi ^4}\left(\frac{17 g_{XR}^2 g_{RR}^3}{4}-\frac{3}{4} g_{XR} g_{XX} g_{RR}^3+3 g_{2L}^2 g_{RR}^3+12 g_{3C}^2 g_{RR}^3 \right. \nonumber\\& \left. +\frac{15 g_{XX}^2 g_{RR}^3}{8}+\frac{17 g_{RR}^5}{4}\right), \\
 \beta_{XX} =& \frac{1}{16 \pi ^2}\left(\frac{53 g_{XR}^2 g_{XX}}{12}-\frac{g_{XR} g_{XX}^2}{4}+\frac{33 g_{XX}^3}{8}\right)+\frac{1}{256 \pi^4}\left(\frac{17 g_{XR}^4 g_{XX}}{4}-\frac{3 g_{XR}^3 g_{XX}^2}{2} \right. \nonumber\\& +3 g_{XR}^2 g_{2L}^2 g_{XX} +12 g_{XR}^2 g_{3C}^2 g_{XX}+\frac{45 g_{XR}^2 g_{XX}^3}{4}+\frac{17}{4} g_{XR}^2 g_{XX} g_{RR}^2 -\frac{9 g_{XR} g_{XX}^4}{4}\nonumber\\ &\left. -\frac{3}{4} g_{XR} g_{XX}^2 g_{RR}^2+\frac{9 g_{2L}^2 g_{XX}^3}{2}  +4 g_{3C}^2 g_{XX}^3+\frac{65 g_{XX}^5}{16}+\frac{15 g_{XX}^3 g_{RR}^2}{8}\right), \\
 \beta_{XR} =& \frac{1}{16 \pi^2}\left(\frac{53 g_{XR}^3}{12}-\frac{g_{XR}^2 g_{XX}}{4}+\frac{33 g_{XR} g_{XX}^2}{8}+\frac{53 g_{XR} g_{RR}^2}{6}-\frac{g_{XX} g_{RR}^2}{4}\right) \nonumber\\ &
 +\frac{1}{256 \pi^4}\left(\frac{17 g_{XR}^5}{4}-\frac{3 g_{XR}^4 g_{XX}}{2}+3 g_{XR}^3 g_{2L}^2+12 g_{XR}^3 g_{3C}^2+\frac{45 g_{XR}^3 g_{XX}^2}{4} \right. \nonumber \\ &+\frac{51 g_{XR}^3 g_{RR}^2}{4}-\frac{9 g_{XR}^2 g_{XX}^3}{4} -3 g_{XR}^2 g_{XX} g_{RR}^2+\frac{9}{2} g_{XR} g_{2L}^2 g_{XX}^2  \nonumber \\ & +6 g_{XR} g_{2L}^2 g_{RR}^2+4 g_{XR} g_{3C}^2 g_{XX}^2+24 g_{XR} g_{3C}^2 g_{RR}^2  +\frac{65 g_{XR} g_{XX}^4}{16} \nonumber \\ & \left. +\frac{105}{8} g_{XR} g_{XX}^2 g_{RR}^2+\frac{17 g_{XR} g_{RR}^4}{2}-\frac{9 g_{XX}^3 g_{RR}^2}{8}-\frac{3 g_{XX} g_{RR}^4}{4}\right). \\
\end{align}

\bibliographystyle{JHEP}
\bibliography{GUT}

\end{document}